\documentclass[12pt]{article}
\usepackage[english]{babel}
\usepackage{amsfonts,amsmath,mathrsfs,stmaryrd,amsthm}
\usepackage{graphicx}
\usepackage{amssymb}
\usepackage{amsmath}
\usepackage{verbatim} 
\usepackage{natbib}
\usepackage{color}
\renewcommand{\hat}{\widehat}
\renewcommand{\tilde}{\widetilde}

\def\bSig\mathbf{\Sigma}

\def\b#1{\mbox{\boldmath $#1$}}
\def\b#1{\mbox{\boldmath $#1$}}

\title{Longitudinal quantile regression in presence of informative drop-out through longitudinal-survival joint modeling}

\author{Alessio Farcomeni\\
Department of Public Health and Infectious Diseases\\
Sapienza - University of Rome \and 
Sara Viviani\\
Department of
Statistics\\
Sapienza - University of Rome}
\date{}
\begin{document}

\label{firstpage}

\maketitle

\begin{abstract}
We propose a joint model for a time-to-event outcome and a quantile 
of a continuous response repeatedly measured over time. 
The quantile and survival processes are associated via shared latent
and manifest variables.
Our joint model provides
a flexible approach to handle informative drop-out in quantile
regression. A general Monte Carlo Expectation Maximization strategy based on
importance sampling is proposed, 
which is directly applicable under any distributional assumption for
the longitudinal outcome and random effects, and parametric and
non-parametric assumptions for the baseline hazard.  
Model properties are illustrated through a 
simulation study and an application to an
original data set about dilated cardiomyopathies. 
\end{abstract}


\section{Introduction}
\label{intro}

In longitudinal studies subjects may be lost to follow-up due to 
events, like death, which are associated with the outcome of
interest. Failure to model drop-out may lead to biased estimates in
such cases. From the reverse perspective, the time trend of a
longitudinal measurement may predict the risk of an event (e.g., a
steadily decreasing CD4 count is predictive of adverse events in HIV
patients). A general account 
of related longitudinal and survival processes can be found in 
\cite{follmann:95}. Participation to a study 
can in general be described by a survival model of time to drop-out.
The simpler way of taking into account informative
drop-out is through pattern-mixture models (e.g.,
\cite{wu:bail:88,wu:bail:89,litt:wang:96}), where the outcome
distribution is specified conditionally on the time to drop-out. 
Selection models on the other hand condition the drop-out mechanism to
unobserved responses directly \citep{digg:kenw:94} or indirectly. 
A simple and effective indirect link between drop-out and unobserved
outcomes is by assuming that these are independent conditionally on 
unobserved shared random effects, as in \cite{wu:carrol:88}. 
One can similarly assume that the risk of event at time
$t$ is influenced by the \textit{expected value} of 
the longitudinal response, as 
in \cite{riz:10} and \cite{rizop:ghosh:11}. In the resulting
Joint Model (JM), the hazard of drop-out is a 
function of the predicted longitudinal outcome, that is, of shared 
random and fixed effects, and
related covariates. See also \cite{litt:95}, 
\cite{hend:et:al:00}, \cite{tsia:davi:04}, and references therein. 

Our concern in this paper is that the expected value of the longitudinal
outcome may not always be the summary of interest. Further, in some
cases it might be difficult to find a suitable transformation to
normality for the outcome, or some resistance to outliers may be
desired. An effective solution to these issues is given by modeling  
conditional quantiles of the longitudinal outcome. Quantile
regression models \citep{koen:05} 
are robust with respect to outliers, so that one can simply model the
median rather than the mean. 
In many biomedical applications 
interest lies furthermore in at least 
one of the tails, and covariates may have different 
effects on different quantiles of a 
distribution.
Examples include longitudinal fetal growth studies, which are usually focused
on low and high quantiles of key anthropometric measurements. 
Finally, quantiles are
invariant to transformations so it is never needed to transform the
outcome. 
%Quantile regression can be applied in different
%areas, see for instance \cite{yu:lu:stan:03}, \cite{mach:mata:05},
%\cite{aust:schu:03}, \cite{cade:noon:flat:05}. 
Longitudinal quantile regression models are proposed among others in 
\cite{koen:04}, who maximizes a penalized version of the likelihood;  
%\cite{koen:04} maximizes a penalized likelihood. 
%\cite{karl:08} derives a weighted approach to accommodate nonlinear
%longitudinal data. 
and \cite{gera:bott:07}, who introduce 
a random intercept and assume the outcome follows 
the asymmetric Laplace distribution (ALD). 
\cite{yuan:bott:09} extend the random intercept model to a 
general linear mixed quantile
regression model. See also \cite{gera:bott:13}.
The ALD assumption is also used in \cite{farc:12}, where 
random effects are time-varying and follow a discrete distribution. 
Informative missing data are ubiquitous in statistical applications,
especially in longitudinal studies, but there are very few approaches 
to quantile regression with informative drop-out.
In \cite{lips:et:al:97} and \cite{yi:he:09} 
estimating equations are weighted proportionally to
the inverse of the probability of drop-out. 
\cite{yuan:yin:10}, in a Bayesian framework, model missingness as a
binary time series sharing a random effect with the quantile
regression process. A common limit of these approaches is that 
drop-out can occur only at one of the observation times of the
longitudinal process. This does not hold in general (e.g., when
measurements are scheduled according to a protocol, and 
death occurs between two visits). Moreover, all approaches proposed so far do
not directly model the strength of association between the
longitudinal and time-to-event processes. The latter is summarized by
non-ignorability parameter(s) in JM as the one we propose. 
A longitudinal quantile regression model 
with ignorable missingness is outlined in \cite{gera:13}. 

In this paper we propose a joint model for 
a right-censored time-to-event outcome and the quantile 
of a continuous response repeatedly measured over time. 
Drop-out is formally defined as a monotone missing pattern, that is,
when the outcome is not measured for a subject, no further
measurements take place for that subject until the end of the study. 

In our approach the quantile and survival processes are associated not
only via shared latent variables or the predicted longitudinal
outcome. It is in fact assumed that the
time-to-event outcome depends on a function, which for simplicity we
assume linear, of both the latent variables and the predicted
quantile of the longitudinal outcome. 
Two non-ignorability parameters are introduced, one for the fixed and
the other for the random part of the linear predictor. 
Our joint model is therefore a flexible approach to handle informative
drop-out in longitudinal quantile regression. 
%From the longitudinal quantile regression standpoint, our joint model provides
%a flexible approach to handle informative drop-out. 
%From the survival standpoint, the measurement error in time dependent
%covariates is avoided by considering the predicted median (or any
%other quantile) rather than the predicted mean. 
%We better discuss below the meaning and implications of this feature of our
%model.
%We suggested different alternatives for the random effects
%distribution. 
%It
%is shown in \cite{rizopoulos:08b} that shared parameter models are
%mostly insensitive to the choice of the random effects distribution as
%the number of occasions grow. This happens mostly since inference
%depends on the {\it posterior} distribution of random effects, which 
%is better and better approximated by a Gaussian density 
%regardless of the prior distribution
%\citep{hsie:06}. We still give different choices  
%since the number of measurement occasions may be low in many
%applications, therefore making it crucial to correctly specify the random
%effects distribution.
Maximum likelihood estimates are efficiently derived by setting up a
Monte Carlo Expectation-Maximization (MCEM) algorithm based on 
importance sampling \citep{douc:et:al:01,levi:case:01}. 
There are two clear advantages of using importance sampling: first,
the resulting MCEM is completely general and straightforward to use
with any distributional assumption for the longitudinal observations
or the random effects. Secondly, it is computationally efficient 
given that we evaluate the posterior distribution only once for each
sample. The MCEM approach,
unlike commonly used quadrature methods, is amenable also to
moderate dimensional random effects.
%In synthesis, our contribution can be summarized as follows: as far as 
%longitudinal quantile regression is concerned, we propose a
%flexible and effective approach to deal with informative drop-out,
%directly modeling the strenght of the association between the drop-out
%and longitudinal process, and allowing for drop-out at any time point
%after the beginning of the study. As far as 
%joint longitudinal-survival models are concerned, we show 
%how to use non-Gaussian
%error distributions (thus for instance modeling conditional quantiles of the
%longitudinal outcome) and non-Gaussian 
%multivariate random effects
%distributions. 
The use of two non-ignorability parameters allows us 
also increase the flexibility of both the
shared parameter model of \cite{wu:carrol:88}, by conditioning the
drop-out process also on the residuals between the predicted longitudinal
outcome and the random effects, and the JM of 
\cite{WULF:TSIAT:97} and \cite{riz:10}, by allowing 
a residual dependence on the random effects. In the first
case, we can say that the two processes are
not only linked by unmeasured heterogeneity, but that their dependence can
also be in part explained through shared observed heterogeneity. 
In the second case, we can say that the
two processes are linked not only via a quantile of 
the longitudinal outcome, but also by a residual unmeasured
heterogeneity. 
%Appropriate constraints, which can be easily tested,
%give back the shared parameter model or the JM. 

The rest of the paper is as follows: in the next section we describe 
the proposed model. In Section \ref{inference} we outline inference, 
we illustrate the approach through simulations in Section \ref{simu},
                           and
in Section \ref{realdata} where we apply
the method to an original data set about patients with
cardiomyopathy. 
Finally, in Section \ref{concl} we conclude with a brief discussion. 

\section{Joint longitudinal quantile and survival regressions}

Let $T_i=\min(T_i^*,C_i)$ denote the observed failure time for
the $i$th individual, $i=1,\ldots,n$, 
taken as the minimum between the true event time $T_i^*$ and the
censoring time $C_i$. Further, let $\Delta_i$ be the corresponding 
event indicator defined by $ \Delta_i = I(T_i^* \leq C_i)$, where
$I(\cdot)$ is the indicator function. 
The continuous outcome $Y_{it}$ is repeatedly observed at times
$t = 1,\ldots,n_i$ before $T_i$,
and is missing for $t \geq T_i$. 
The longitudinal outcome at observation times is collected in $\b y_i = \{y_i(t) : t \leq T_i\}$.
We assume that the 
longitudinal process is associated with $T_i^*$, i.e with the true
event time, but, as customary in survival analysis, 
is independent of the censoring time $C_i$. 

We let $\b X_{it}$ denote a vector of predictors used to model only
the longitudinal outcome, $\b H_{it}$ a vector of shared predictors, 
and $\b W_i$ a vector of (time fixed) predictors used to model only 
the survival process. These are associated with fixed effects
$\b\beta$, $\b\delta$ and $\b\gamma$, respectively. We 
then have a vector $\b Z_{it}$ of predictors associated with 
$q$ dimensional random effects $\b u_i$, and two non-ignorability
parameters $\alpha_1$ (associated with fixed effects) and $\alpha_2$
(associated with random effects). 
Our model can be expressed by a set of two equations, 
one for the longitudinal and the other one for the survival outcome:
\begin{eqnarray}
\label{jqsm}
\begin{cases}
y_{it} = \b\beta' \b X_{it} + \b\delta' \b H_{it} + \b u_i'\b Z_{it} +\epsilon_{it} = \tilde{\tau}_{it} +  \epsilon_{it} \\
h(T_i | \mathcal{T}_{iT_i}, \b W_i ; \b \gamma, \alpha_1, \alpha_2) =
h_0(T_i)\exp\{\b \gamma' \b W_i + \alpha_1  \b\delta'\b H_{iT_i} +
\alpha_2\b u_i'\b Z_{iT_i}  \} ,
\end{cases}
\end{eqnarray}
where the first equation gives the longitudinal model and the second
the time-to-event model, and $h_0(s)$ is a baseline function. 
Specifically, the model for the longitudinal outcome $\tilde{\tau}_{it}$
is formulated along the usual lines for mixed effects models
\citep{verb:mole:00} and the model for the time-to-event outcome is
based on the subject-specific hazard function $h(T_i)$
\citep{cox:72,and:82}. 
The risk of drop-out is conditional on $\mathcal{T}_{iT_i} =\{\tilde{\tau}_{iu}: 0\leq u \leq T_i\}$,
i.e. the error-free longitudinal process history up to time $T_i$.
The model is completed by a distributional assumption for  
the shared latent distribution, that is, by specifying 
$\b u_i \sim f(\b u_i)$. Few options are discussed in 
Section \ref{ranef}.

The degree of dependence between the longitudinal and the survival
processes is measured by the association parameters $\alpha_1$ and $\alpha_2$, 
which are introduced to assess potential non-ignorability of the
missing data mechanism. 
In doing so, we admit two sources of non-ignorability: a part that can
be explained through observed heterogeneity in $\b H_{it}$ (but not
$\b X_{it}$) and a part that is due to unobserved heterogeneity. 
The log-hazard ratios associated with $\b H_{it}$ can be estimated as
$\alpha_1\b\delta$, while those associated with $\b W_i$ are directly
estimated as $\b\gamma$. All parameters are identifiable even if we
multiply some of them in the survival model equation. 

It is worth noting how our proposed model (\ref{jqsm}) generalizes
previous work. If we fix $\alpha=\alpha_1=\alpha_2$, $\beta=0$, and assume a
Gaussian distribution for the error, we obtain a usual formulation for
the JM. Otherwise, with an ALD error distribution (see below), we obtain a joint
quantile regression model. Here,
 the link between the time-to-event and longitudinal
processes is summarized by the $\alpha (\b\delta'\b H_{it} + \b u_i \b
Z_{it}) = \alpha\tilde{\tau_{it}}$ term in the
model for the hazard function. 
If $\alpha_1 \neq \alpha_2$ and $\beta=0$,  
$$
\alpha_1  \b\delta' \b H_{it} + \alpha_2 \b u_i'\b Z_{it} = \alpha_1
\tilde{\tau_{it}} + (\alpha_2-\alpha_1) \b u_i'\b Z_{it}, 
$$
that is, we generalize JM models by allowing for a residual dependence
on unobserved heterogeneity as summarized by the shared random
effects. When $\b\beta \neq 0$, we further assume that some predictors
may be related only to the longitudinal outcome but may not contribute
to explain non-ignorability. 
Similarly, a shared parameter model is obtained if we fix $\alpha_1=0$. 
Specifically, if $Z_{it}=(1\ t)$ and we have Gaussian error terms and
random effects, we exactly obtain the classical model in \cite{WULF:TSIAT:97}. 
When $\alpha_1>0$ we condition the survival process also on the difference between the predicted longitudinal
outcome and the random effects. It is straightforward to fully
generalize (\ref{jqsm}) by including also random effects 
that are not shared in each part of the model, and by letting $W_i$ be
time-dependent. In this way we would have, in each model, 
separate and shared covariates both for random and fixed effects. 
We do not pursue this explicitely to keep notation
simple, but we note that our MCEM strategy can be directly adapted to
this completely general case. 

We start describing each part of the model separately, then we outline how they
are linked by obtaining the observed likelihood. 

\subsection{The longitudinal model}

The parametric assumption on the error distribution of the
longitudinal outcome drives our target for inference. 
If we assume that $\varepsilon_{it}$ follows a zero-centered Gaussian,
we work with the conditional expectation of the outcome.
On the other hand, if we assume an ALD, $\tilde{\tau}_{it}$ does not represent the conditional mean of
$Y_{it}$ anymore, but its conditional $0 <\tau< 1$ quantile. Note that
$\tau$ is pre-specified and fixed. The resulting density of $Y_{it}$,
conditionally on covariates and random effects, is given by  
\begin{equation}
\label{ald}
f(Y_{it}|\b X_{it},\b H_{it},\b Z_{it},\b \beta,\b \delta,\b u_i,\sigma) =
\frac{\tau(1-\tau)}{\sigma}
\exp\left\{-\rho\left(\frac{Y_{it}-\b\beta'\b X_{it} -\b\delta'\b
  H_{it}-\b u_i'\b Z_{it}}{\sigma}\right)\right\},
\end{equation}
where $\rho(u)=u\{\tau - I(u<0)\}$ is the quantile loss
function and $\sigma>0$ is a scale parameter.
When $Z_{it}=1$, we obtain a random intercept model
\citep{gera:bott:07}. The ALD is justifiable 
since maximum likelihood is exactly equivalent to
minimization of the quantile loss function, when no parametric assumptions
are made on $\varepsilon$ \citep{yu:moye:01}. Further, it can be seen 
in simulation studies to lead to good estimates even when the
residuals are not ALD distributed (see for instance
\cite{yuan:bott:09}, \cite{farc:12}). 

\subsection{The survival model}

The time-varying baseline risk function $h_0(T_i)$ in (\ref{jqsm}) can 
be seen as the risk obtained when all covariates and random effects
are exactly zero. We may wish to specify a flexible parametric
form for $h_0(t)$ (e.g., $h_0(t)=\eta t^{\eta-1}$, leading to a Weibull model), as in for instance 
\cite{rizop:ghosh:11}, \cite{riz:12}, \cite{vivi:alfo:rizo:13} or 
we may leave it unspecified as in \cite{follmann:95}. The inferential
strategy for obtaining the MLE is slightly different in the two cases,
as we discuss in the next section. 

The time-to-event distribution can in both cases be written as
\begin{eqnarray}
\label{p.t}
f(T_i, \Delta_i \mid \b u_i)&=&
f(T_i \mid \mathcal{T}_{it},\b W_i)^{\Delta_i}S(T_i \mid
\mathcal{T}_{it}, \b W_i)^{1-\Delta_i} \nonumber\\
&=&h(T_i \mid \mathcal{T}_{it},\b W_i)^{\Delta_i}S(T_i \mid
\mathcal{T}_{it},\b W_i)  ,
\end{eqnarray}
where $S(\cdot)$ denotes the survival function, i.e.
$$
S(T_i \mid \mathcal{T}_{it}, \b W_i) = \exp\left\{ - \int_0^{T_i}
h_0(s)\exp\{\b\gamma' \b W_i + \alpha_1  \b\delta' \b H_{is} +
\alpha_2\b u_i'\b Z_{is}  \} ds 
\right \} 
$$
while $h(T_i \mid \mathcal{T}_{it},\b W_i)$ is given by the second equation in (\ref{jqsm}). 

\subsection{The random effects model}
\label{ranef}

A commonly used distribution for random effects is the multivariate
normal. This is convenient to work with when a Gaussian assumption is
formulated also for the longitudinal outcome. The multivariate
normal may not be satisfactory anyway when the number of occasions is
small (\cite{rizopoulos:08b},\cite{hsie:06}) 
and /or the dimensionality $q$ of $\b Z_{it}$ is large. Further, we may
often expect a slower convergence of the posterior distribution of the
random effects to a multivariate normal when modeling lower or upper quantiles. 
Two valid alternatives for the random effects
distribution are given by a multivariate $T$ with $k$ degrees of freedom: 
$$
f(\b u_i|\b\Sigma) \propto |\b\Sigma|^{-1/2}\left(1+\frac{1}{k}
\b u_i'\b\Sigma^{-1}\b u_i\right)^{-\frac{k+q}{2}}, 
$$
which can be used to capture fat tails of the random effects; 
and a multivariate ALD, which is often suggested in the quantile
regression framework \citep{yuan:bott:09}: 
$$
f(u_i|\b\Sigma) \propto |\b\Sigma|^{-1/2}\left(\frac{\b
  u_i'\b\Sigma^{-1}\b u_i}{2}\right)^{v/2} K_v\left(\sqrt{\b
  u_i'\b\Sigma^{-1}\b u_i}\right)
$$
where $v=(2-q)/2$ and $K_v(\cdot)$ is the modified Bessel function of
the third kind.
The most appropriate random effects distribution may be chosen 
for instance using the Bayesian Information Criterion (BIC) or the
Akaike Information Criterion (AIC), see for instance \cite{verblesaf:96}.

\subsection{Observed likelihood}
\label{sec:lik}

In what follows, $\b\theta$ is a short-hand notation for model
parameters, that is, $\b\beta$, $\b\delta$, $\alpha_1$, $\alpha_2$,
$\b\gamma$, $\sigma$, $\b\Sigma$ and any parameter associated with baseline hazard $h_0(s)$. 
The joint likelihood contribution of the longitudinal and survival processes for the $i$-th
subject is obtained integrating the conditional distributions in
(\ref{jqsm}) over the random effects space: 
\begin{eqnarray}
\label{indlik}
f(T_i, \Delta_i,\b Y_i;\b\theta) = \int f(\b Y_i | \b u_i) f(T_i,
\Delta_i |\b u_i) f(\b u_i| \b\Sigma) d \b u_i,
\end{eqnarray}
where $f(T_i, \Delta_i \mid \b u_i)$ is given in equation (\ref{p.t}) and
\begin{eqnarray*}
f(\b Y_i |\b u_i) = \prod_{t=1}^{n_i} f(Y_{it} |\b u_i),
\end{eqnarray*}
while $f(Y_{it}  \mid
\b u_i)$ is given in equation (\ref{ald}).
The observed data log-likelihood for the joint quantile regression
model is then 
\begin{eqnarray}
\label{lik}
\ell(\b\theta) = \sum_i \log f(T_i, \Delta_i, y_i;\b\theta).
\end{eqnarray}

The integrals involved in (\ref{indlik}) 
are usually tackled in the JM context through quadrature methods (see
for instance \cite{riz:12}, \cite{vivi:alfo:rizo:13}). While effective in one or
two dimensions, quadrature methods 
tend to become too slow or less precise as the dimensionality of the random effect
distribution grows.  
Furthermore, quadrature methods should have to be tailored to the random effects distribution (e.g., a 
Gauss-Hermite quadrature would be best for Gaussian random effects,
while a Gauss-Laguerre may be better under other assumptions).
In next section we propose a Monte Carlo strategy which is completely general,
and allows us to set up an algorithm which is easily adapted to any assumption on the
random effects and to any functional form for the two parts of the
JM. 

\section{Estimation of the proposed model}
\label{inference}

We propose a MCEM algorithm for fitting the proposed model. This 
involves alternating two steps until convergence: (i) sampling from the
posterior distribution for the random effects, given the data and
current values of the parameters (Monte Carlo step), and obtaining the
conditional expected value of the complete data log likelihood (E-step),
(ii) maximizing the latter (M-step). 
The algorithm is guaranteed to converge to a local optimum. In order
to increase the odds of obtaining the global optimum, we perform a
multistart. The first run starts from estimates
obtained from separate longitudinal and survival models, which are readily available. 
Other runs are initialized by randomly perturbing the 
deterministic initial solution. 
A second remark regards how to obtain standard errors and confidence intervals.
When performing quantile regression on the longitudinal outcome, 
we use a non-parametric bootstrap strategy 
(\cite{buch:95}, \cite{andr:buch:00}). 
We preserve the dependency structure in the data by resampling 
subjects rather than separately resampling the 
outcomes (and related predictors). 
Under the usual regularity conditions, tests on the regression 
parameters may then be simply performed by using Wald statistics based 
on the standard errors. 

The MCEM algorithm is based on the complete likelihood, that
is, the likelihood we would have if we could observe the random effects. 
The individual contribution to the complete data log-likelihood can be obtained as 
\begin{equation}
\label{IndcomplLik}
\log f(T_i, \Delta_i, \b Y_i,\b u_i;\b\theta) = \log f(T_i, \Delta_i
|\b u_i;\b \theta) + \log f(\b Y_i |\b u_i;\b \theta)+ \log f(\b u_i|\b\Sigma).
\end{equation}
The MCEM algorithm is completely general and can be simply adapted to any
distributional assumption for the longitudinal error and random
effects distribution. 
Assuming an ALD for the error distribution of the longitudinal
measurements, the complete data log-likelihood is as follows:
\begin{eqnarray}
\label{complLik}
\ell_c(\b\theta)&=& \sum_i \log f(T_i, \Delta_i,\b Y_i,\b u_i;\b\theta)=  \\ \nonumber
%&=& \sum_i \log f(T_i, \Delta_i , Y_i | u_i; \theta) + \log f(u_i | T_i, \Delta_i, Y_i; \theta)
&=& \sum_i \log f(Y_i | \b u_i;\b \theta) + \sum_i \log f(T_i,
\Delta_i |\b u_i;\b \theta) + \sum_i \log f(\b u_i|\b\Sigma) \\ \nonumber
&=&  -\log \sigma\sum_i n_i  - \sum_i\sum_{t=1}^{n_i}
\rho\left(\frac{Y_{it} - \b\beta'\b X_{it} - \b\delta'\b H_{it}-\b
  u_i'\b Z_{it}}{\sigma}\right) \\ \nonumber
&+&  \sum_i \Delta_i \log h_0(T_i) + \sum_i \Delta_i\b\gamma'\b W_i + \alpha_1
\sum_i \Delta_i \b\delta'\b H_{iT_i} + \alpha_2 \sum_i \Delta_i\b
u_i'\b Z_{iT_i}\\ \nonumber
&-&  \sum_i \int_0^{T_i} h_0(s) \exp\{\b\gamma'\b W_i + \alpha_1 \b \delta'\b H_{is}
+ \alpha_2\b u_i'\b Z_{is}\} ds \\\nonumber
&+& \sum_i \log f(\b u_i|\b\Sigma). 
\end{eqnarray}

\subsection{Monte Carlo E-step}
\label{e}

The conditional expected value of (\ref{IndcomplLik}) 
for the $i$-th subject at the $j$-th iteration of the algorithm is
expressed as 
\begin{eqnarray}
\label{Expect}
\mathbb{E}[ \ell_c(\b\theta\mid  T_i, \Delta_i, \b Y_i,\b u_i) ] &= \sum_i \int  &[
\log f\left(\b Y_i | \b u_i;\b\theta\right) + \log f\left( T_i, \Delta_i |\b u_i; \b\theta\right)  \nonumber \\ 
&+& \log f\left(\b u_i; \b\theta\right)
] 
f\left(\b u_i \mid  T_i, \Delta_i,\b Y_i;\b \theta^{(j)}\right) d\b u_i ,
\end{eqnarray}
where $\b \theta^{(j)}$ denotes the current value of the parameters. 
The posterior distribution of the random effects is 
 \begin{eqnarray} 
\label{post.b}
f(\b u_i \mid  T_i, \Delta_i,\b Y_i;\b \theta^{(j)}) &\propto & f(\b
Y_i, T_i, \Delta_i,\b u_i;\b \theta^{(j)}) \nonumber\\
&=&  f(T_i, \Delta_i |\b u_i;\b \theta^{(j)})  f(\b Y_i |\b u_i;\b \theta^{(j)})
f(\b u_i |\b \Sigma^{(j)}).
\end{eqnarray}
Straightforward algebra can be used to see that (\ref{post.b})
simplifies to 
\begin{eqnarray*}
f(\b u_i \mid  T_i, \Delta_i,\b Y_i;\b \theta^{(j)}) &\propto & 
\exp\left\{-\rho\left(\frac{Y_{it}-\b\beta^{(j)'}\b X_{it}
  -\b\delta^{(j)'}\b H_{it}-
  \b u_i'\b Z_{it}}{\sigma^{(j)}}\right)+ \Delta_i\alpha_2^{(j)}\b
u_i'\b Z_{iT_i}\right\}\\
&\phantom{a}& \exp\left\{ -\int_{0}^{T_i} h_0^{(j)}(s)
\exp\{\alpha_2^{(j)}\b u_i'\b Z_{is}\}\right\} f(u_i |\b \Sigma^{(j)})
\end{eqnarray*}

In order to work with (\ref{Expect}), we need to marginalize the joint
distribution with respect to the
multivariate random effect posterior distribution. The resulting integral is conveniently approximated 
through Importance Sampling (IS).  Importance sampling proceeds by obtaining a random 
sample $(v_{i1}, \ldots, v_{im_i^{(j)}})$ from a proposal distribution
$g(\cdot)$. 
The IS identity 
$$
\int \ell_c(\b\theta)  f(\b u_i \mid  T_i,
\Delta_i,\b Y_i;\b \theta^{(j)}) d\b u_i = 
\int \ell_c(\b\theta)  \frac{f(\b u_i \mid  T_i,
\Delta_i,\b Y_i;\b \theta^{(j)})}{g(\b u_i)} g(\b u_i) d\b u_i
$$
is used to approximate (\ref{Expect}). More formally, the expression in (\ref{Expect}) is approximated as 
\begin{eqnarray}
\label{ExpApp}
\mathbb{E}[ \ell_c(\b\theta) ]  &\approx& \sum_i \sum_{b=1}^{m_i^{(j)}} [
\log f\left(\b Y_i |\b v_{ib};\b \theta^{(j)}\right) + \log f\left( T_i,
\Delta_i |\b v_{ib};\b \theta^{(j)}\right) \\
\nonumber 
&\phantom{+}& + \log f\left(\b v_{ib} \mid \b \theta^{(j)}\right)] w_{ib}, 
\end{eqnarray}
where 
$$
\tilde{w}_{ib} = \frac{f(\b v_{ib} \mid  T_i,\Delta_i,\b Y_i;\b
  \theta^{(j)})}{g(\b v_{ib})}
$$
and 
$
w_{ib} = \frac{\tilde{w}_{ib}}{\sum_b \tilde{w}_{ib}}. 
$
%In words, the complete likelihood expected value is 
%then simply obtained through the sum over $i$ of (\ref{ExpApp}), as observations across
%subjects are conditionally independent. 
In this work we proceed as in \cite{levi:case:01}, where
we sample $(v_{i1}, \ldots, v_{im_i^{(j)}})$ from the posterior
distribution at the initial parameter estimates using adaptive
rejection Metropolis sampling \citep{gilks:95}, and then update the
weights at each
iteration. The MC sample size $m_i^{(j)}$ 
is increased to control the MC error \citep{levi:case:01,eichoff:04}. 
Finally, the observed likelihood (\ref{lik}) can be
directly approximated as $\prod_i \sum_b \tilde{w}_{ib}$. 
The latter is used to check convergence of the MCEM and after 
convergence for testing and computation of information criteria. 
In summary, the E-step is given by updating of
$\tilde{w}_{ib}$, $w_{ib}$, and evaluation of the likelihood. 

\subsection{M-step}
\label{m}

We outline here the M-step, which consists in maximizing the
approximated conditional expected value of the complete likelihood
with respect to $\b\theta$. 

When $h_0(s)$ is left completely unspecified, 
we obtain a Nelson-Aalen \citep{nels:72,aale:78} type estimator as 
\begin{equation}
\label{hath0}
\hat h_0(s) = \sum_{i=1}^n \frac{\Delta_i I(T_i=s)}{\sum\limits_{i:
 T_i \geq s} \sum_{b=1}^{m_i^{(j)}} w_{ib}\exp\{\b\gamma'\b W_i +
  \alpha_1 \b\delta' \b H_{is} + \alpha_2\b v_{ib} '\b Z_{is}\}},
\end{equation}
by setting to zero the score equations for $h_0(s)$
\citep{WULF:TSIAT:97}. See also \cite{niel:gill:ande:sore:92} and
\cite{gill:92}. The
expression (\ref{hath0}) is not an explicit solution as it
depends on other parameters. Nevertheless, it can be plugged in the
conditional expected value of the complete likelihood, 
thus obtaining a profile complete likelihood (complete with respect to
the random effects, and profiled with respect to the non-parametric
baseline along the lines of \cite{cox:72}). 
If instead we specify a
parametric form for $h_0(s)$, e.g, $h_0(s)=\eta s^{\eta-1}$, we can
plug-in this expression and update any parameter involved in $h_0(s)$
within the rest of the M-step.

When a Gaussian distribution is assumed for the longitudinal
measurements, regression coefficients and hazard ratios are updated
through a one-step Newton-Raphson algorithm (which is easily adapted
from \cite{WULF:TSIAT:97}), while the variance of the random term 
$\epsilon_{it}$ in (\ref{jqsm}) is estimated through the usual closed
form expression. 

When we assume an ALD for the longitudinal measurements, the M-step is
complicated by the presence of the quantile check function. 
An estimator of $\sigma$, dependent on 
the other parameters, can be explicitely obtained as: 
\begin{equation}
\label{hatsigma}
\hat \sigma = \frac{1}{\sum_i n_i} \sum_{i=1}^n\sum_{t=1}^{n_i} 
\sum_{b=1}^{m_i^{(j)}} w_{ib}\rho\left(y_{it}-\b\beta'\b X_{it} -\b\delta'
  \b H_{it}-\b v_{ib}'\b Z_{it}\right). 
\end{equation} 
The vector of regression parameters and
dispersion parameter for the ALD is block updated using one step of the \cite{neld:mead:65}
numerical optimization algorithm, majorizing (\ref{ExpApp}) after 
plug-in of (\ref{hatsigma}) and (\ref{hath0}) or the parametric
formula of $h_0(s)$.  
The resulting expected complete likelihood is 
\begin{eqnarray*}
\sum_i \Delta_i \log({\hat h}_0(T_i))+\sum_{i=1}^n 
\Delta_i\b \gamma'\b W_i + \alpha_1 \sum_{i=1}^n 
\Delta_i \b\delta'\b H_{iT_i} + \alpha_2 \sum_{i=1}^n
\Delta_i\sum_{b=1}^{m_i^{(j)}} w_{ib}\b v_{ib}'\b Z_{iT_i}&\phantom{+}&\\
-\sum_{i=1}^n 
\sum_{b=1}^{m_i^{(j)}}w_{ib}\int_0^{T_i} \hat h_0(s)\exp\{\b\gamma'\b W_i +
\alpha_1 \b\delta' \b H_{is} + \alpha_2 \b v_{ib}'\b Z_{is}\}\ ds&\phantom{+}& \\
- \sum_{i=1}^n\sum_{t=1}^{n_i} 
\sum_{b=1}^{m_i^{(j)}}w_{ib} \rho\left(\frac{\b Y_{it} - \b\beta'\b X_{it} - \b\delta'
  \b H_{it}- \b v_{ib}'\b Z_{it}}{\hat\sigma}\right)
-\log(\hat\sigma)\sum_{i=1}^n n_i&\phantom{+}&.
\end{eqnarray*}
The integral involved in the expression above
(and similarly in any expression where $f(T_i,\Delta_i|\b v_{ib};
\b\theta)$ appears also at the E-step) reduce to sums when a
non-parametric baseline is used, and can instead be approximated using
one-dimensional Gauss-Kronrod quadrature (e.g.,
\cite{kaha:mole:nash:89}) when a parametric assumption
is formulated for $h_0(s)$. 

The M-step is concluded by maximizing the approximated conditional
expected value of the complete likelihood with respect to parameters
involved in the distribution of the random effects. This is readily
accomplished under any of the assumptions we have proposed in Section
\ref{ranef} using the method of moments. In all cases in fact 
$\b\Sigma$ can be updated as the weighted
empirical covariance matrix of the random effects sampled at the
E-step. 

\section{Simulations}
\label{simu}

In this section we illustrate our approach through a simulation study. We
evaluate the bias and standard deviation of the estimates for our
proposed model for data Missing Not At Random (MNAR), 
and compare with a model which ignores informative
drop-out (Missing At Random - MAR model). 

For $n=\{250,500\}$, $\alpha_1=\{0,1\}$, $\alpha_2=\{0,1\}$, $\tau=\{0.25,0.5,0.75\}$ we
fix $\b\beta=\b\delta=\b\gamma=(1\ 1)$ and $\sigma=1$. We assume random
effects arise from a centered bivariate normal distribution with
standard deviations equal to 0.3 and correlation 0.16.
%We generate $n$ bivariate normal random effects. 
We let $\b Z_{it}=(1\ t)$, $\b H_{it}=(h_{i1}\ h_{i2}*t)$, $\b
X_i=(1\ x_i)$; with $h_{i1}$, $h_{i2}$, $x_i$, $W_{i1}$ and
$W_{i2}$ generated from
independent standard normals. By 
also fixing $h_0(s)=1$ it is possible to exactly obtain 
the survival distribution as
$$
S(t|\b u_i,\b H_i,\b W_i) = \exp\left\{
-\frac{e^{\alpha_1(\delta_1H_{i1}+\delta_2H_{i2t})+\alpha_2(u_{i1}+u_{i2}t)+
 \b\gamma'\b W_i}-e^{\alpha_1\delta_1H_{i1}+\alpha_2u_{i1}+
  \b\gamma'\b W_i}}{\alpha_2u_{i2}+\alpha_1\delta_2h_i}\right\}
$$
when $\alpha_1 \neq 0$ or $\alpha_2 \neq 0$ and 
$$
S(t|U,H,W) = \exp\{-te^{\b\gamma'\b W_i}\}
$$
when $\alpha_1=\alpha_2=0$. 
The expression above can be inverted to obtain $T_i$ after generating
$n$ random variates uniformly distributed on the unit interval. We then 
let the censoring time 
$C_i/5$ be distributed according to a Beta with parameters 4 and 1, in
order to obtain a censoring proportion around 25\%. 
We allow for a maximum of six observation times for each subject, 
at $t=0, 1/4, 1/2, 3/4, 1, 3$. Longitudinal
observations before drop-out 
are independently generated from an ALD for the $\tau$-th
quantile, centered on 
$$
\b\beta'\b X_i+\b\delta'\b H_{it}+\b u'\b Z_{it},
$$
and with dispersion parameter $\sigma$.
We fit our joint model with parametric baseline distribution
and two separate models (one for the longitudinal process and one for
the time-to-event, therefore obtaining MAR estimates), 
based on each generated data set. 

For each setting we report
the bias and standard deviation of the estimates averaged over
$B=1000$ replicates, and further averaged over groups of parameters
for $\b\beta$, $\b\delta$ and $\b\gamma$. Results are shown 
in Table \ref{simures}, where it can be seen that the MNAR model 
has a very low
bias and standard deviation of the estimates for all values of
$\alpha_1$ and $\alpha_2$, with very few exceptions which are likely
due to random fluctuation. Results are consistent
across all quantiles, with a slightly larger MSE for
quantiles distant from the median, as expected. 

\begin{table}[ht]
\centering
\caption{Bias and standard deviation of the estimates of the proposed
  model on simulated
  data for different values of $n$, $\alpha_1$ and $\alpha_2$. Results
are based on $B=1000$ replicates.}
\label{simures}
\begin{tabular}{rrrcccccccccc}
  \hline
\multicolumn{13}{c}{$\tau=0.25$} \\
&&& \multicolumn2c{$\b\beta$} & \multicolumn2c{$\b\delta$} &
  \multicolumn2c{$\b\gamma$} & \multicolumn2c{$\alpha_1$} & \multicolumn2c{$\alpha_2$} \\
$n$ & $\alpha_1$ & $\alpha_2$ & bias & s.d. & bias & s.d & bias & s.d & bias & s.d. & bias & s.d \\
  \hline
250 & 0 & 0 & -0.004 & 0.107 & 0.001 & 0.124 & 0.020 & 0.085 & -0.001 & 0.044 & -0.002 & 0.131\\
250 & 0 & 1 & -0.013 & 0.104 & -0.002 & 0.119 & -0.023 & 0.100 & -0.001 & 0.066 & -0.076 & 0.177\\
250 & 1 & 0 & -0.003 & 0.109 & 0.009 & 0.094 & 0.020 & 0.088 & 0.021 & 0.101 & -0.008 & 0.146\\
250 & 1 & 1 & -0.012 & 0.105 & -0.003 & 0.108 & -0.021 & 0.106 & -0.025 & 0.108 & -0.068 & 0.189\\
500 & 0 & 0 & -0.002 & 0.074 & -0.000 & 0.082 & 0.006 & 0.058 & -0.001 & 0.031 & -0.003 & 0.091\\
500 & 0 & 1 & -0.011 & 0.072 & 0.002 & 0.082 & -0.034 & 0.072 & 0.001 & 0.046 & -0.076 & 0.120\\
500 & 1 & 0 & -0.003 & 0.074 & 0.004 & 0.061 & 0.008 & 0.061 & 0.008 & 0.067 & -0.006 & 0.092 \\
500 & 1 & 1 & -0.011 & 0.074 & -0.004 & 0.075 & -0.033 & 0.073 & -0.038 & 0.076 & -0.063 & 0.118\\
\hline 
\multicolumn{13}{c}{$\tau=0.5$} \\
&&& \multicolumn2c{$\b\beta$} & \multicolumn2c{$\b\delta$} &
  \multicolumn2c{$\b\gamma$} & \multicolumn2c{$\alpha_1$} & \multicolumn2c{$\alpha_2$} \\
$n$ & $\alpha_1$ & $\alpha_2$ & bias & s.d. & bias & s.d & bias & s.d & bias & s.d. & bias & s.d \\
\hline
250 & 0 & 0 & -0.002 & 0.096 & -0.000 & 0.113 & 0.020 & 0.084 & -0.001 & 0.044 & -0.000 & 0.126 \\
 250 & 0 & 1 & -0.014 & 0.092 & -0.003 & 0.107 & -0.020 & 0.100 & -0.001 & 0.066 & -0.064 & 0.164 \\
 250 & 1 & 0 & -0.001 & 0.097 & 0.005 & 0.087 & 0.020 & 0.087 & 0.022 & 0.096 & -0.006 & 0.138 \\
 250 & 1 & 1 & -0.015 & 0.094 & -0.005 & 0.101 & -0.019 & 0.105 & -0.030 & 0.107 & -0.074 & 0.175 \\
 500 & 0 & 0 & 0.001 & 0.067 & -0.000 & 0.076 & 0.005 & 0.057 & -0.001 & 0.031 & -0.004 & 0.083 \\
 500 & 0 & 1 & -0.008 & 0.063 & 0.000 & 0.074 & -0.028 & 0.071 & -0.001 & 0.045 & -0.060 & 0.106 \\
 500 & 1 & 0 & -0.000 & 0.066 & 0.001 & 0.058 & 0.006 & 0.059 & 0.010 & 0.064 & -0.003 & 0.088 \\
 500 & 1 & 1 & -0.008 & 0.066 & -0.006 & 0.069 & -0.031 & 0.072 & -0.035 & 0.072 & -0.056 & 0.119\\
\hline
\multicolumn{13}{c}{$\tau=0.75$} \\
&&& \multicolumn2c{$\b\beta$} & \multicolumn2c{$\b\delta$} &
  \multicolumn2c{$\b\gamma$} & \multicolumn2c{$\alpha_1$} & \multicolumn2c{$\alpha_2$} \\
$n$ & $\alpha_1$ & $\alpha_2$ & bias & s.d. & bias & s.d & bias & s.d & bias & s.d. & bias & s.d \\
\hline  
 250 & 0 & 0 & -0.001 & 0.108 & 0.002 & 0.122 & 0.019 & 0.085 & -0.003 & 0.044 & -0.007 & 0.134 \\
 250 & 0 & 1 & -0.013 & 0.104 & -0.003 & 0.119 & -0.023 & 0.100 & -0.001 & 0.066 & -0.076 & 0.169 \\
 250 & 1 & 0 & 0.003 & 0.111 & 0.004 & 0.091 & 0.021 & 0.088 & 0.024 & 0.099 & -0.002 & 0.145 \\
 250 & 1 & 1 & -0.012 & 0.107 & -0.006 & 0.107 & -0.020 & 0.106 & -0.028 & 0.112 & -0.079 & 0.178 \\
 500 & 0 & 0 & 0.004 & 0.075 & -0.003 & 0.083 & 0.006 & 0.058 & -0.001 & 0.031 & -0.004 & 0.088 \\
 500 & 0 & 1 & -0.006 & 0.070 & -0.000 & 0.080 & -0.030 & 0.072 & -0.001 & 0.046 & -0.066 & 0.109 \\
 500 & 1 & 0 & 0.004 & 0.075 & -0.002 & 0.062 & 0.007 & 0.060 & 0.014 & 0.068 & -0.004 & 0.092 \\
 500 & 1 & 1 & -0.007 & 0.073 & -0.009 & 0.075 & -0.032 & 0.073 & -0.033 & 0.076 & -0.062 & 0.120\\
   \hline
\end{tabular}
\end{table}

In Table \ref{simures2} 
we report the ratio of the bias and the variance of the estimates of
the MAR over the MNAR model, based on the average bias for all
parameters, separately for the longitudinal and survival parameters. 
These ratios are close to the unity when $\alpha_1=\alpha_2=0$, with our model
generally performing slightly better given that the MNAR model 
assumes $\b\delta$ parameters are equal in the longitudinal and
survival parts. When $\alpha_1$ or
$\alpha_2$ are non-zero, the ratios of the variances of the estimates
are still close to the unity, but the ratios of biases increase
substantially. The bias of the MAR model may be up to 30 times the
bias of our joint model. The effect of $\alpha_1$
is often stronger than the effect of $\alpha_2$, but this is likely
only due to the fact that in all simulated settings there is a 
larger heterogeneity due to the shared covariates with
respect to the unobserved heterogeneity due to random effects. 
As could be expected, the ratios are generally increasing with $n$, 
given that the MSE of the joint model is infinitesimal.

\begin{table}[ht]
\centering
\caption{Ratios of bias and variance of the estimates obtained with
  the MAR model ($bias_{MAR}$, $var_{MAR}$) and with our model
  ($bias$, $var$) for the longitudinal $(Y)$ and survival $(T)$
  part of the model. Results are shown for different values of $n$,
  $\alpha_1$ and $\alpha_2$ and are based on $B=1000$ replicates.}
\label{simures2}
\begin{tabular}{rrrcccc}
  \hline
\multicolumn{7}{c}{$\tau=0.25$}\\
$n$ & $\alpha_1$ & $\alpha_2$ &
  $\left|\frac{bias_{MAR}(Y)}{bias(Y)}\right|$ &
  $\frac{var_{MAR}(Y)}{var(Y)}$ & $\left|\frac{bias_{MAR}(T)}{bias(T)}\right|$ &
  $\frac{var_{MAR}(T)}{var(T)}$ \\
  \hline
250 & 0 & 0 & 1.548 & 1.203 & 0.800 & 1.115 \\ 
250 & 0 & 1 & 6.616 & 1.241 & 1.266 & 1.315 \\ 
250 & 1 & 0 & 2.876 & 1.573 & 9.990 & 0.929 \\ 
250 & 1 & 1 & 5.868 & 1.385 & 8.051 & 0.814 \\ 
500 & 0 & 0 & 1.264 & 1.252 & 0.802 & 1.197 \\ 
500 & 0 & 1 & 4.214 & 1.270 & 1.954 & 1.287 \\ 
500 & 1 & 0 & 6.966 & 1.654 & 5.954 & 0.996 \\ 
500 & 1 & 1 & 7.227 & 1.412 & 24.926 & 0.834 \\ 
  \hline
\multicolumn{7}{c}{$\tau=0.5$}\\
$n$ & $\alpha_1$ & $\alpha_2$ &
  $\left|\frac{bias_{MAR}(Y)}{bias(Y)}\right|$ &
  $\frac{var_{MAR}(Y)}{var(Y)}$ & $\left|\frac{bias_{MAR}(T)}{bias(T)}\right|$ &
  $\frac{var_{MAR}(T)}{var(T)}$ \\
  \hline
250 & 0 & 0 & 1.091 & 1.121 & 0.909 & 1.129 \\ 
 250 & 0 & 1 & 1.679 & 1.204 & 1.464 & 1.333 \\ 
 250 & 1 & 0 & 5.659 & 1.439 & 7.579 & 0.948 \\ 
 250 & 1 & 1 & 4.646 & 1.289 & 10.572 & 0.804 \\ 
 500 & 0 & 0 & 0.865 & 1.188 & 0.930 & 1.221 \\ 
 500 & 0 & 1 & 6.936 & 1.221 & 1.208 & 1.339 \\ 
 500 & 1 & 0 & 2.756 & 1.501 & 6.758 & 1.041 \\ 
 500 & 1 & 1 & 5.797 & 1.290 & 29.194 & 0.851 \\ 
  \hline
\multicolumn{7}{c}{$\tau=0.75$}\\
$n$ & $\alpha_1$ & $\alpha_2$ &
  $\left|\frac{bias_{MAR}(Y)}{bias(Y)}\right|$ &
  $\frac{var_{MAR}(Y)}{var(Y)}$ & $\left|\frac{bias_{MAR}(T)}{bias(T)}\right|$ &
  $\frac{var_{MAR}(T)}{var(T)}$ \\
  \hline
 250 & 0 & 0 & 1.308 & 1.223 & 1.010 & 1.151 \\ 
 250 & 0 & 1 & 4.972 & 1.280 & 1.283 & 1.318 \\ 
 250 & 1 & 0 & 3.895 & 1.586 & 7.153 & 0.934 \\ 
 250 & 1 & 1 & 4.191 & 1.367 & 10.346 & 0.783 \\ 
 500 & 0 & 0 & 1.866 & 1.255 & 1.304 & 1.201 \\ 
 500 & 0 & 1 & 6.733 & 1.276 & 1.135 & 1.303 \\ 
 500 & 1 & 0 & 4.370 & 1.672 & 5.414 & 1.014 \\ 
 500 & 1 & 1 & 7.216 & 1.347 & 27.891 & 0.836 \\ 
   \hline
\end{tabular}
\end{table}

\section{Application to dilated cardiomyopathy data}
\label{realdata}

In this section we briefly illustrate the proposed approach on an
original data set about patients with dilated cardiomyopathy. 
Data refers to $n=659$ consecutive patients who begun treatment for dilated
cardiomyopathy in the cardiovascular department of ``Ospedali
Riuniti'' in Trieste, Italy. Patients were
enrolled at first treatment and scheduled for follow-up after 6
months, 1, 2, 3, 4, 6 and 10 years, with only 25\% of the patients
having complete records. Maximum follow-up time before
cardiovascular death or censoring due to loss at follow-up or transplant 
is 25 years, with a total of 212 events (32\%). 

Dilation of the left ventricular is known to lead to hearth
failure, and in many cases an ethiological basis cannot be
identified \citep{merlo:11}. The goal of this study is to compare 
patients with mild dilation of the left ventricular (Mildly
Dilated CardioMyopathy or MDCM) with respect to patients with a general dilation of unrecognized
ethiology (Idiopatic Dilated CardioMyopathy or IDCM). MDCM patients
are generally believed to be at a slightly lower risk (e.g., \cite{kere:90}), but the
physiological reasons are still unrecognized. 

Our longitudinal outcome of interest is the left ventricular 
ejection fraction (LVEF), that is, the volumetric fraction of blood 
pumped out of the ventricle with each heartbeat. Note that LVEF is a
bounded outcome, but all measurements are far from the boundaries 
so that we unlike \cite{bott:cai:mcke:10} we can avoid any
transformations. Dropout occurs due to cardiovascular death, 
and it can be easily expected that the two
processes are related as patients with a lower ejection fraction are
at higher risk of death. For example, a univariate Cox model for the
baseline LVEF gives an hazard ratio (HR) of 0.95 for each percentage point,
with $p<1e-16$. Furthermore, LVEF is skewed and its skewness seems to change
over time. This leads to two issues: first, using a classical joint
model after transformation of the LVEF would be awkward, as the
optimal transformation is different at each time point. Secondly,
modeling the mean of the transformed LVEF would not be as meaningful
from a clinical perspective than directly modeling quantiles,
which is also straightforward to interpret. We mostly are interested in low
quantiles (like the 10th or the 15th), in the terziles or quartiles
for the outcome. 
See for instance \cite{sand:03}, \cite{Clements01062005}, \cite{ndre:etal:07}, 
\cite{Cowie03102012}. Consequently, we explore 
the ejection fraction distribution by evaluating 
$\tau=0.1,0.15,0.25,0.33,0.5,0.66,0.75$. 

We model the longitudinal outcome conditionally on an intercept and
age at baseline ($\b X$ matrix), on the indicator of MDCM and its
interaction with time ($\b H$ matrix). Besides covariates in the $\b
H$ matrix, we let the hazard of death depend on gender (1 for males)
and indicator
of New York Hearth Association (NYHA) functional classes
I or II at baseline ($\b W$ matrix). For more details 
on NYHA functional classes see for instance \cite{merlo:11} and
references therein. We also include a
shared subject-specific random intercept and a random slope, that is,
$\b Z=(1\ t)$. Given that the number of follow-up times is slightly large, 
we use a normality assumption for $\b u_i$. We also have compared with
a multivariate $T$ and multivariate Laplace, with analogous results
which we do not report for reasons of space. We only mention that the
multivariate normal distribution is chosen using AIC and BIC
criteria. We estimate our proposed
model both with a parametric Weibull baseline and with a
non-parametric baseline. Likelihood, AIC, and BIC at each quantile are
reported in Table \ref{aicandbic}. 

\begin{table}[ht]
\caption{Log-likelihood at convergence, AIC and BIC for our model fit 
with a parametric and non-parametric baseline, at
  different quantiles of interest, on the dilated cardiomyopathy data.}
\label{aicandbic}
\fbox{
\begin{tabular}{ccccccc}
   & \multicolumn3c{non-parametric baseline} & \multicolumn3c{parametric baseline} \\
$\tau$ & $\ell(\b\theta)$	& BIC	& AIC	& $\ell(\b\theta)$&	BIC	&AIC\\
.1&	-10515.214&	21064.255&	21054.428&	-10602.111&	21240.867&	21228.222\\
.15	&-10449.858&	20933.543&	20923.716&	-10461.581&	20959.807&	20947.162\\
.25&	-10289.721&	20613.269&	20603.442&	-10322.538&	20681.721&	20669.076\\
.33	&-10197.313&	20428.453&	20418.626&	-10235.107&	20506.859&	20494.214\\
.5	&-10154.227&	20342.281&	20332.454&	-10175.175&	20386.995&	20374.350\\
.66	&-10223.996&	20481.819&	20471.992&	-10229.257&	20495.159&	20482.514\\
.75&	-10300.913&	20635.653&	20625.826&	-10308.303&	20653.251&	20640.606\\
\end{tabular}}
\end{table}

Based on those results we select the non-parametric
baseline for all quantiles. We compare the estimates 
with those obtained under a MAR model in Table \ref{mdcm_res}. 

\begin{table}[ht]
\caption{Estimates for the MNAR and MAR models for the MDCM data. An
  asterisk indicates that estimates in the column are significant
  at the 5\% level for all quantiles $\tau$. The models are based on a non-parametric baseline
  for the survival process and two dimensional shared random effects.}
\label{mdcm_res}\fbox{
\begin{tabular}{ccccccccc}
& \multicolumn8c{MNAR model} \\
 & \multicolumn4c{Longitudinal outcome} &
  \multicolumn4c{Survival outcome}\\ 
$\tau$ & Int* & Age* & MDCM* & MDCM:time* & Gender* & NYHA* & $\alpha_1$* &
  $\alpha_2$* \\
.1&	22.535& -0.038&  7.754& -0.038&	0.538 &-0.912&   -0.044 & -0.042\\
.15&	25.134 &-0.048&  7.768 &-0.028	&0.527 &-0.895&  -0.037 & -0.057\\
.25&	29.255 &-0.068 & 7.752 &-0.042	&0.593 &-0.990&  -0.017 & -0.068\\
.33&	32.813 &-0.095 & 7.752 &-0.018	&0.592 &-0.988&  -0.018 & -0.087\\
.5&	39.086 &-0.124 & 7.749 &-0.093	&0.542 &-0.918&  -0.031 & -0.063\\
.66&	44.437 &-0.148 & 7.734 &-0.136	&0.566 &-0.952&  -0.011 & -0.097\\
.75&	47.355 &-0.156 & 7.744 &-0.044	&0.591 &-0.987&  -0.021 & -0.094\\
\hline
& \multicolumn8c{MAR model} \\
 & \multicolumn4c{Longitudinal outcome} &
  \multicolumn4c{Survival outcome}\\ 
& Int* & Age* & MDCM* & MDCM:time* & Gender* & NYHA* & $\alpha_1$ &
  $\alpha_2$\\
.1&	29.700 &-0.141& 7.580& 0.041& 0.600& -1.096 & 0.000 & 0.000\\
.15&	36.266 &-0.251& 8.027& 0.076&0.600 &-1.096 &0.000& 0.000\\
.25&	35.175 &-0.130& 5.262& 0.011&0.600 & -1.096 &0.000& 0.000\\
.33&	35.246 &-0.109& 6.516& 0.068&0.600 & -1.096 &0.000& 0.000\\
.5&	40.048 &-0.137& 8.289& 0.028&0.600 &-1.096 &0.000& 0.000\\
.66&	44.541 &-0.123 &5.711& 0.022&0.600 &-1.096 & 0.000& 0.000\\
.75&	44.284 &-0.088& 5.558& 0.017&0.600 & -1.096 &0.000& 0.000\\
\end{tabular}}
\end{table}

There is a stronger and stronger effect of age on LVEF
as $\tau$ increases, while the effect of MDCM is slightly constant
with $\tau$, with a negative interaction with time. 
Males are at slightly higher risk of death and patients in
lower functional NYHA classifications are at a lower risk. After
adjusting for these covariates, the significant and negative estimates
for $\hat\alpha_1$ lead us to conclude that MDCM is an {\it
  independent} predictor of a slightly lower risk of death, even after
considering its effect on LVEF. We could expect negative estimates for
$\alpha_1$ and $\alpha_2$ as longitudinal measurements 
and survival time are positively dependent. 
%Note that estimates for $\hat\alpha_1$ and $\hat\alpha_2$ are 
%rather different, hence a classical joint model would probably have been restri%ctive for the data at hand. 
%Since $\hat\alpha_1$ is significant, also 
%classical shared-parameter models would have been restrictive. 

Ignoring drop-out may lead to an important bias. First of all, the
intercepts estimated with the MAR models are slightly larger than
those obtained with the MNAR models for all $\tau$, except
$\tau=75\%$. This is in line with the expected consequences of
drop-out in the low quantiles of the longitudinal outcome
distribution. Secondly, under the MAR models a significant {\it
  positive} interaction between MDCM and time is estimated. This may
be due to the fact that 
subjects with higher LVEF tend to drop-out later and to be in the MDCM
class more often, resulting in a positive bias when ignoring the
informative drop-out. We conclude by noting that given the results in
Table \ref{mdcm_res} we can conclude there is some sensitivity to drop-out
for the data at hand within the proposed class of models. As clearly
noted in \cite{mole:et:al:08} one can never test the MAR assumption. 

\section{Conclusions}
\label{concl}

Informative drop-out may bias parameter estimates 
both in mean and quantile regression if ignored. 
As our data example suggests, the problem may be
stronger for quantiles corresponding to a higher rate of events. 
In our example we have checked that sensitivity to drop-out
is milder in high quantiles than in low quantiles, for instance. 
Moreover, the brief simulation study reported confirms that, as long
as the informative drop-out process is ignored, bias of the parameter
estimates may be substantial and, more importantly, may not decrease
with the sample size. 

The proposed approach allows to simultaneously model the quantile 
of a longitudinal outcome and the hazard of drop-out, allowing them to
share part of the observed and unobserved heterogeneity. 
Our model can be applied with right-censored event times occuring
between two scheduled visits, when
drop-out times coincide with observation times for the longitudinal
process, and also when the observation times are not scheduled in
advance. We have generalized shared-parameters and
joint-models in different directions: first of all, we have proposed
an alternative parametric assumption for the longitudinal error, the
ALD, which allows to fit quantile regression models. Secondly, we have
proposed two alternative random effects distributions.  
A general efficient MCEM strategy has been 
used to fit our model under any of those assumptions. 

In our example we have specified different values for the quantile of
interest $\tau$ for illustration. It can be argued that under
conditional independence assumptions the total likelihood is the sum
of the likelihood based on each quantile, hence this approach is equivalent to
simultaneously fitting the model for different values of $\tau$, and
$\tau$-specific parameters. When more than one quantile is of interest
in applications, one could also allow dependence (e.g., over the
random effects at each quantile) or $\tau$-homogeneity (e.g., for the
variance of the random effects). Model estimation under these
assumptions is at the moment grounds for further work. 

\section*{Acknowledgements}
The authors are grateful to the Cardiovascular department of
``Ospedali riuniti'', Trento, Italy, and in particular to Giulia
Barbati for permission to use the dilated cardiomyopathy data. 

\bibliography{biblio}
\bibliographystyle{metron}

\label{lastpage}
\end{document}